\documentclass[aps,pra,twocolumn,superscriptaddress,showpacs,longbibliography]{revtex4}
\usepackage{graphicx}
\usepackage{natbib}
\usepackage[breaklinks]{hyperref}
\usepackage{amsmath}
\usepackage{mathtools}

\begin{document}

%\preprint{}

\title{Operating a near-concentric cavity at the last stable resonance}
\author{ChiHuan Nguyen}
\affiliation{Centre for Quantum Technologies, 3 Science Drive 2, Singapore 117543}
\author{Adrian Nugraha Utama}
\affiliation{Centre for Quantum Technologies, 3 Science Drive 2, Singapore 117543}
\author{Nick Lewty}
\affiliation{Centre for Quantum Technologies, 3 Science Drive 2, Singapore 117543}
\author{Christian Kurtsiefer}
\affiliation{Centre for Quantum Technologies, 3 Science Drive 2, Singapore 117543}
\affiliation{Department of Physics, National University of Singapore, 2 Science Drive 3, Singapore 117542}
%\thanks{}
\email[]{christian.kurtsiefer@gmail.com}
\date{\today}

\begin{abstract}
Near-concentric optical cavities of spherical mirrors can provide technical
advantages over the conventional near-planar cavities in applications
requiring strong atom-light interaction, as they concentrate light in a very
small region of space.
However, such cavities barely support stable optical modes, and thus impose
practical challenges. Here, we present an experiment
where we maintain a near-concentric cavity at its last resonant length for
laser light at 780\,nm resonant with an atomic transition. 
At this point, the spacing of two spherical mirror surfaces is $207(13)$\,nm
shorter than the critical concentric point, corresponding to a stability
parameter $g=-0.999962(2)$ and a cavity beam waist of $2.4\,\mu$m. 
\end{abstract}

% insert suggested PACS numbers in braces on next line

\maketitle
\section{Introduction}
Optical cavities are widely used, ranging from lasers and gravitational wave
detectors to experiments in quantum physics exploring nonlinear atom-light
interaction.
In particular, atom-cavity systems with ultra-high finesse cavities are a key
component in
demonstrations of quantum logic gates, distributed quantum networks, quantum
metrology, and sensing applications
\citep{Reiserer2015,Ritter2012,Reiserer2014} using cavity quantum
electrodynamics. 
The intricate high-reflectivity coatings of the cavity mirrors used in these
experiments, however, can pose a challenge on scaling systems up.  
Therefore, new types of optical cavities and resonance structures to enhance
the electrical field of an optical mode have been considered recently \citep{Cox2018,Nguyen2017}.
One such a cavity design that has been experimentally demonstrated is a
near-concentric Fabry-Perot cavity \citep{Morin1994,Nguyen2017}. Outside the
field of cavity QED, these near-unstable cavities have been considered to
reduce the influence of thermal noise of the mirror coatings on gravitational wave
detectors \citep{Wang2018}.

Near-concentric cavities are formed by two spherical mirrors with a
normal separation $l_{cav}$ just short of the sum of the two radii of
curvatures. 
Among all geometries of Fabry-Perot cavities, near-concentric
cavitiesexhibit the tightest focus of the cavity modes. A tight focus
leads to a large electrical field in the focus, and therefore a strong coupling
to an atom trapped there.

 In a near-concentric cavity with a length of several millimeters, the
 effective mode volume can be very small and comparable to state-of-the-art
 cavities of micrometer lengths. The relatively large mirror separation
 permits to form a cavity with a narrow spectral linewidth already with
 mirrors of low finesse that are less challenging to make. Other advantages
 are a better optical access to the focal region, which can be helpful for
 preparation and manipulation of quantum emitters like atoms or ions.  

 Furthermore, the near-degeneracy in resonant frequencies of transverse modes of near-concentric cavities is an intriguing feature to explore the physics of multi-mode strong coupling in cavity quantum electrodynamics \citep{Ballantine2017}.

However, near-concentric cavities have not been widely
explored yet, mainly because they require mirrors that cover a relatively
large solid angle, and because of technical hurdles of stabilizing both the
longitudinal and transverse mirror positions.

Here, we report on a compact design of a symmetric near-concentric cavity
with a length of 11\,mm corresponding to a free spectral range of 13.6\,GHz,
and the strategy to stabilize it to the last few stable 
resonances near the concentric point.
The design is intended to study atom-light interaction but can be easily adapted to a wider range of experiments.
\begin{figure}
  \includegraphics[width=0.8\columnwidth]{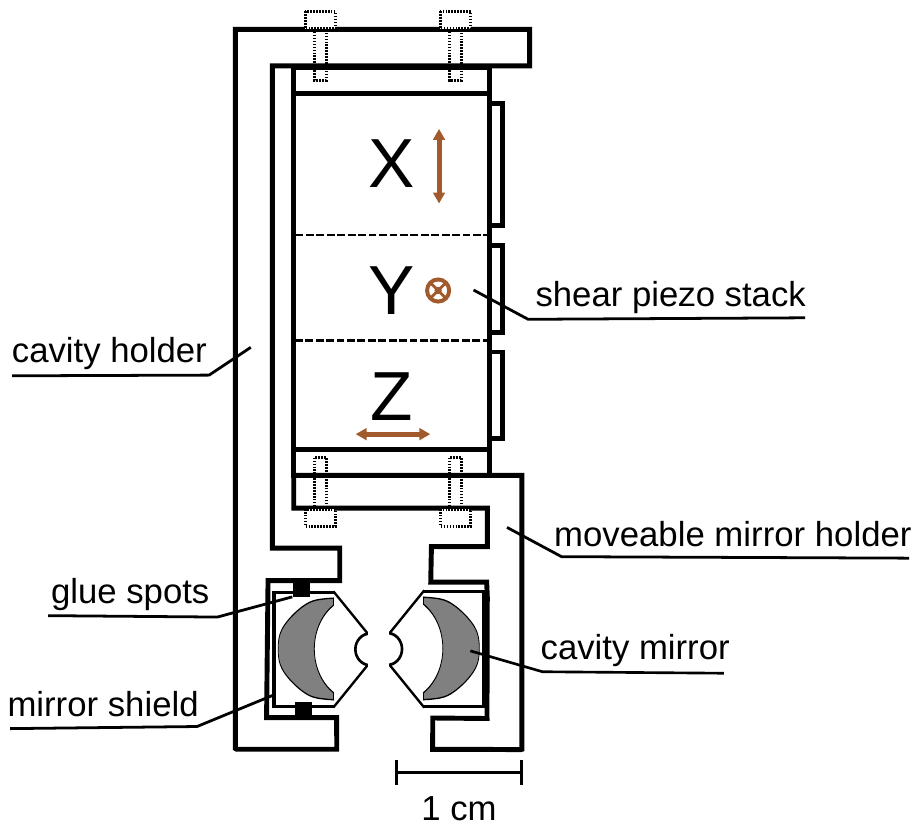}
  \caption{\label{fig:cms}
  Schematic of a near-concentric cavity assembly. Arrows indicate the moving
  directions of piezo segments. 
}
\end{figure}

\section{Optical setup}
\subsection{Cavity design}
The spherical cavity mirrors of our cavity have a radius of curvature $R_{C}= 5.5$\,mm, and a nominal reflectivity of $R=99.5 \%$ at a wavelength of 780\,nm.
 For effective matching to an external probe mode, we use
 an ellipsoidal second surface to 
 transform a collimated (Gaussian) input mode into a spherical wave at the
 mirror surface.  The details of characterization and abberations analysis of
 the mirrors can be found in \citep{Durak2014}.

 To align the cavity and correct for thermal drifts, we place one of the
 cavity mirrors on a shear piezo stack % (PI P-153.6710H),
 with a travel range
 of $\pm \,5\, \mu$m in three orthogonal directions.
The cavity mounting system is shown in Fig.~\ref{fig:cms} and fits into a
cuvette of a vacuum chamber which provides convenient optical 
access to the cavity focus for other optical beams preparing atoms in
experiment.Except for the cavity mirror shields, all the mechanical parts are made from Titanium to reduce the structural change of the mounting system due to thermal fluctuation.

\subsection{Alignment procedure}
 The relatively large numerical aperture of near-concentric cavity modes and
the aspheric outside surface of the cavity mirrors require that the optical
axes of the two cavity mirrors coincide -- a requirement that is much less
critical in conventional cavity arrangements. Additionally, the absolute
transverse separation of the mirror surfaces needs to be near the critical
distance within the moving range of the piezo translator. 

 A collimated laser beam between two fiber couplers defines a reference line
 for the alignment of the cavity mirrors. One cavity mirror is pre-assembled
 in the movable mirror holder. Then the other cavity mirror
is gradually moved into the cavity holder on an external translation stage. 
Throughout the alignment process, the reflected beams from the two cavity
mirrors are monitored and ensured to couple back to the optical fibers. This
keeps the tilt of the mirrors under control, and provides a coarse transverse
alignment between the two cavity mirrors. The fine adjustment is carried out
by a piezo system on the external translation stage,
 before the mirror is glued into the aligned position inside the cavity holder. 
\begin{figure*} [ht!]
\centering
  \includegraphics[width=\textwidth]{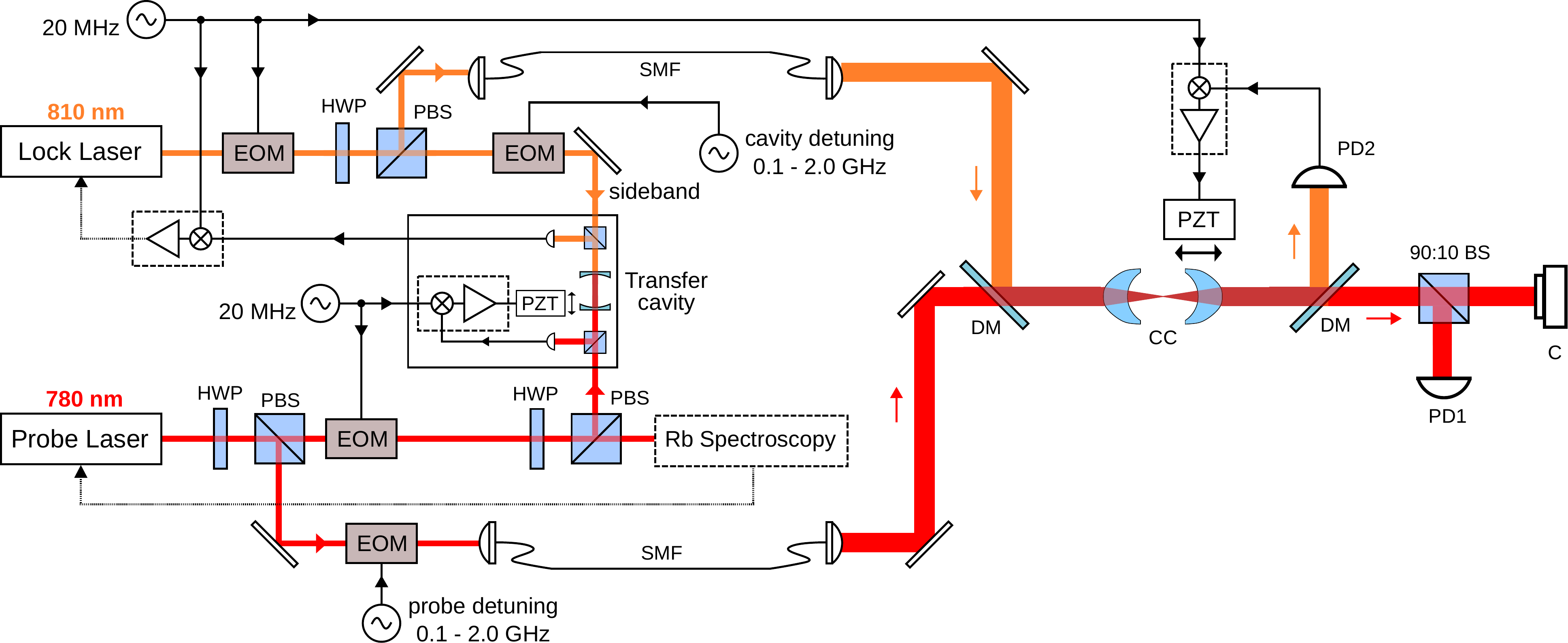}
  \caption{\label{fig:opticalsetup}
Locking scheme of the near-concentric cavity setup. Red and orange lines indicate the beams from 780 nm probe laser and 810 nm lock laser, respectively. The frequency of the probe laser is stabilized to a D2 transition of \textsuperscript{87}Rb by modulation transfer spectroscopy. The lock laser's sideband is locked to resonance of the transfer cavity, which in turn is stabilized to the probe laser. 
The frequency of the lock laser can be tuned by adjusting the sideband frequency. The near-concentric cavity is stabilized to the lock laser. All cavity locking schemes use the standard Pound-Drever-Hall technique with 20 MHz phase modulation. The cavity transmission of probe and lock lasers are separated by a dichroic mirror (DM). A camera with linear response (C) and a photodetector (PD1) are placed at the cavity transmission's 780 nm arm to observe the resonant modes. PBS: polarization beam splitter. BS: beam splitter. SMF: single-mode fibers.
}
\end{figure*}
\subsection{Longitudinal locking scheme}
As we intend to use the cavity for cavity QED experiments, we need to
stabilize its resonance frequency with respect to an atomic transition
independently from the light used to interact with the atom-cavity
system. Therefore, a separate wavelength, far detuned from the atomic
transition under consideration is used.
 The optical layout of the locking scheme is shown in
 Fig.~\ref{fig:opticalsetup}. Laser light at wavelengths of 780\,nm and 810 \,nm
 is coupled into the near-concentric cavity, which we refer to as probe and
 the lock light, respectively. The probe laser is referenced to a D2
 transition of \textsuperscript{87}Rb via a modulation transfer
 spectroscopy~\cite{McCarron2008}. The stability of the probe laser is passed
 to the lock laser at 810\,nm wavelength via a transfer cavity. For that, the
 transfer cavity is first locked to the probe laser. Then, one sideband
 generated by an electro-optical-modulator (EOM) on the 810\,nm light is
 locked to a resonance of the transfer cavity. By tuning the frequency of the
 sideband, the frequency of the lock laser can be adjusted, and is chosen such
 that the probe and lock beams are simultaneously resonant with the
 near-concentric cavity. The probe light itself can be tuned around the atomic
 resonance through another EOM in a similar way.
 All locks use the standard Pound-Drever-Hall technique~\citep{Drever1983}
 with additional sidebands at 20\,MHz which never reach the near-concentric cavity.

\section{Cavity length measurement}
The eigenmodes of an optical resonator with spherical mirrors can be described
by Laguerre-Gaussian (LG) functions, as they form a complete basis to
solutions of the paraxial wave equation, and capture well the cylindrical
symmetry of the 
resonator along the optical axis~\cite{Allen1992}. We denote cavity modes as
$\textrm{LG}_{nlp}$ with integer number mode indices $n,l,p$. Modes of
different $n$ identify longitudinal modes, while $l$ and $p$
characterize the transverse mode profile.
The resonance frequencies of the cavity modes are fixed by the condition that the round-trip phase shift in the cavity must be an integer multiple of $2 \pi$.
As the cavity length approaches concentric point, the shift of the transverse mode frequencies approaches the free spectral range. Therefore all transverse modes become co-resonant in the concentric regime.

Making use of this property, we determine the cavity length by measuring the
spacing of resonant frequencies between the fundamental mode
$\textrm{LG}_{00}$  and the transverse mode $\textrm{LG}_{10}$. Under paraxial
approximation, the resonance frequencies of the cavity with identical
spherical mirrors are given by
\begin{equation}\label{Eq:res_freq}
\nu_{n,l,p}=n\frac{c}{2l_\textrm{cav}}+ \left(1+|l|+2p\right)\frac{c}{2l_\textrm{cav}} \frac{\Delta \psi}{\pi}\,,
\end{equation}
where $c$ is the speed of light, $\Delta \psi=2\tan^{-1}\left(l_{cav}/{2z_{0}}\right)$ the Gouy phase difference after one round trip of $\textrm{LG}_{00}$, and $z_{0}$ the Rayleigh range of the cavity \citep{Saleh2001}.
\begin{figure} %[ht]
\centering
  \includegraphics[width=\columnwidth]{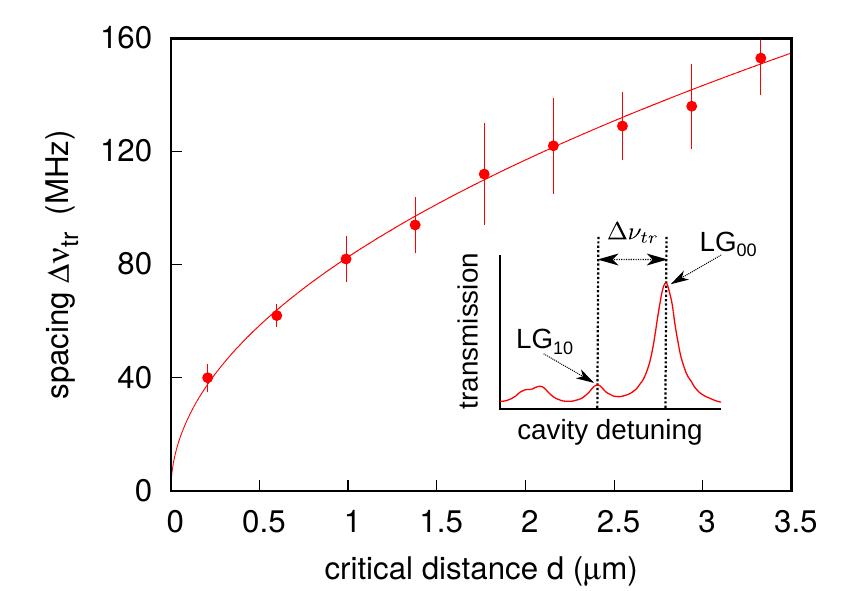}
  \caption{\label{fig:tms}
 Transverse-mode frequency spacing ($\Delta\nu_{tr}$) at different critical distance ($d$) of cavity lengths that are resonant with 780\,nm laser. The solid line is the fit based on Eq.~\ref{eq:tms}. Error bars show the standard deviation of the measurement. The inset shows a typical cavity transmission spectrum and the derived $\Delta\nu_{tr}$.
}
\end{figure}
From Eq.~(\ref{Eq:res_freq}) follows an expression for frequency spacing of $\textrm{LG}_{00}$ and $\textrm{LG}_{10}$ in terms of $l_{cav}$ and $R_{C}$,
\begin{equation}
\Delta \nu_\textrm{tr} = \nu_{n00}-\nu_{n10} =\frac{c}{2l_\textrm{cav}} \left(1- \frac{\cos^{-1}g}{\pi}  \right)\,,\label{eq:tms}
\end{equation}
where $g=1-l_{cav}/R_{C}$ is the stability parameter.

In the experiment, we obtain cavity transmission spectra by varying the cavity
length within a free spectral range. We record spectra at different resonant
cavity lengths, and apply a peak detection algorithm to determine the resonant
frequencies. Different transverse modes are disinguished by imaging the
intensity distribution of light transmitted through the cavity with a
camers. The frequency measurements are calibrated with a frequency marker
obtained by modulating the probe laser with an electro-optical phase modulator
(EOM). 

Figure~\ref{fig:tms} shows the transverse mode frequency spacing at different cavity lengths which are resonant with the 780 nm laser. We define the critical distance as $d=2R_{C}-l_{cav}$.
From a fit of experimental data points to Eq.~\ref{eq:tms}, we determine  $d= 207(13)$ nm at the last stable resonance, which corresponds to the stability parameter $g=-0.99996(2)$. This is consistent with our observation that when increasing the cavity length by another half wavelength, the cavity enters the unstable regime and exhibits lossy cavity modes (see Fig.~\ref{fig:cavity_trans} and the next section).

The good agreement between the experimental data, including the last resonant
point, and the fit based on the paraxial equation prompts us to discuss the
validity of the paraxial approximation in our near-concentric cavity. In the
paraxial approximation, the complex electric field amplitude of a beam
propagating in the $z$ direction can described as
$E\left(x,y,z\right)=u\left(x,y,z\right)e^{-ikz}$, where $k$ is the
longitudinal wave vector component, and $u(x,y,z)$ an envelope function; its
slow variation in the paraxial approximation requires
\begin{equation}
\left| \frac{\partial^2 u}{\partial^2 z}  \right| \ll \left|2k \frac{\partial u}{\partial z}\right|\,.\label{eq:field}
\end{equation}
Conventionally, Eq.~\ref{eq:field} is considered valid for optical beam
components with an angle with the optical axis up to $\approx$30
degrees~\cite{siegman86}. Transverse fundamental near-concentric cavity modes
(LG$_{n00}$) have a beam divergence of $\theta = {\lambda}/{\pi w_{0}}$, where
$\lambda$ is the wavelength of the resonant mode (780\,nm in our case) and
$w_{0}$ is the cavity beam waist. Taking the beam divergence now as a
characteristic angle with the optical axis, the divergence limit of 30 degrees
for the paraxial approximation corresponds to $w_{0} \leq 496$\, nm, or
equivalently $d \leq 0.5 $\,nm. The region of critical distances we explore is
much larger, so the paraxial approximation is still valid.
% to describe our near-concentric cavity.
Note that the definition of the critical distance $d$ and validity of
Eq.~(\ref{Eq:res_freq}) are based on a meaningful definition of a mirror
surface position. The thickness of the dielectric Bragg stacks forming the
mirrors for our cavity exceeds by far the critical distances $d$ for the last
stable longitudinal resonances, so the absolute position of the mirror surface
has to refer to an effective position of these Bragg stacks.

\section{Cavity mode analysis}
\begin{figure} [ht!]
\centering
  \includegraphics[width=\columnwidth]{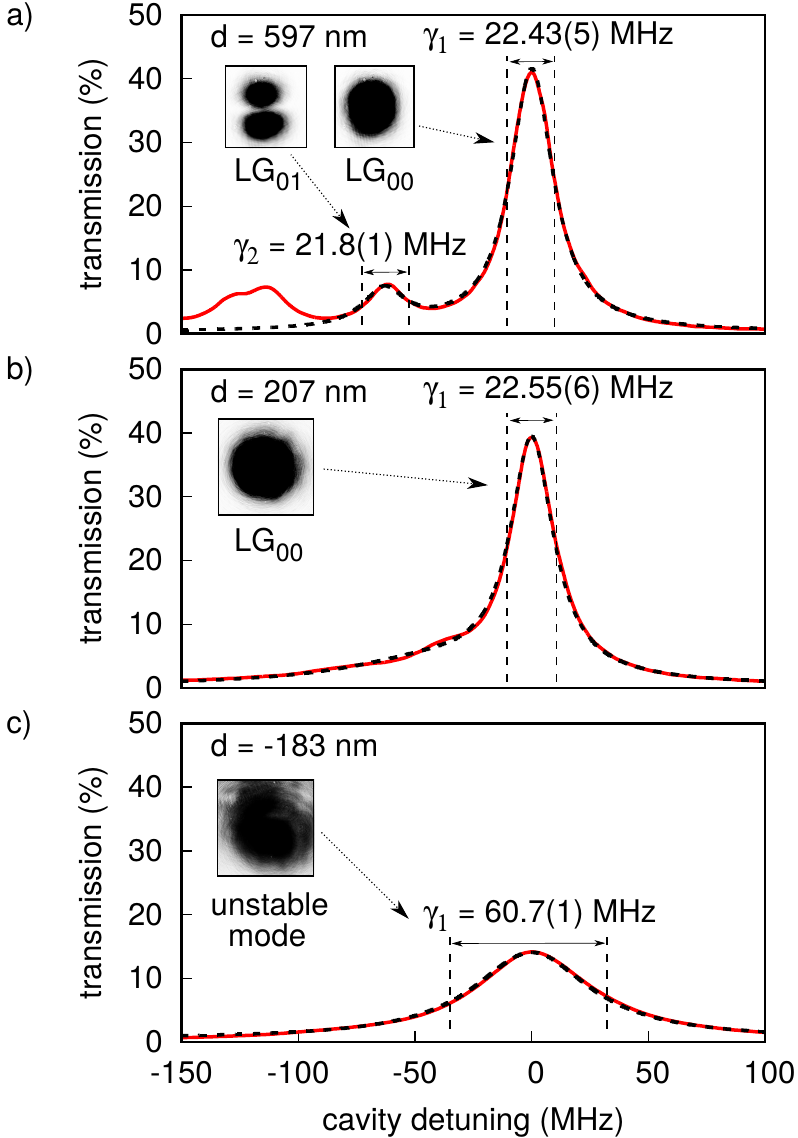}
  \caption{\label{fig:cavity_trans}
 Cavity transmission spectra, measured by detuning the cavity through small
 changes (few nm) in the cavity length. (a) $d= 
 597\,$\,nm. The dashed line is a fit to a sum of two Lorentzian
 functions modeling two resonant peaks. (b) $d=
 207\,$\,nm. Transverse modes become degenerate and form a long tail extending
 out to the lower frequencies. (c) $d=-183\,$\,nm. The cavity is in the
 unstable regime. Insets show the transmitted transverse mode profiles
 recorded by a (partly saturated) camera.
}
\end{figure}

Earlier observations indicated that the cavity finesse reduces significantly
as the cavity is pushed toward the geometrical instability
regime~\cite{Haase2006}. In contrast to this, possibly due to refined 
manufacturing techniques of large angle spherical mirror surfaces, we find
that our near-concentric cavity can maintain the transmission and linewidth at
the last two resonant cavity lengths before the unstable regime. Typical
cavity transmission spectra are shown in Fig.~\ref{fig:cavity_trans}. To
characterize line widths of sligthly overlapping cavity modes, we model the
cavity transmission by a sum of two Lorentzian functions,
 \begin{equation}
 T(\nu)=\frac{T_{1}}{4(\nu-\nu_{1})^2/\gamma_{1}^2+1}+\frac{T_{2}}{4(\nu-\nu_{2})^2/\gamma_{2}^2+1}\,,\label{eq:transmission}
 \end{equation}
 where $T_{1(2)}$ are transmission coefficients, $\nu_{1(2)}$ resonant
 frequencies, and $\gamma_{1(2)}$  the line widths of cavity modes
 $\textrm{LG}_{00}$ and $\textrm{LG}_{10}$.
 From a fit of Eq.~(\ref{eq:transmission}) to cavity transmission spectra we
 then determine the cavity parameters at multiple cavity lengths.

We find that at $d=207$\,nm, the last longitudinal resonance of the cavity,
fundamental mode maintains the cavity linewidth and transmissions of other
longitudinal resonances.  The observed
linewidth of the fundamental mode $\textrm{LG}_{00}$ still agrees well with the
nominal value of 21.7 MHz determined from the cavity mirror's design
reflectivity of 0.995 at a wavelength of 780\,nm. In contrast, the transverse
modes start to overlap at the last resonant length, and the probe laser
simultaneously couples to multiple cavity modes such that the second cavity
mode becomes difficult toidentify, resulting in a broadened effective
linewidth %, which is determined from the fit to be 
of 98(2) MHz from the fit. An increase of the cavity length by another half
wavelength leads to a decrease in the cavity transmission and, an increase in
the cavity linewidth, which are indicative of unstable cavity modes.

Besides the scattering and absorption loss, due to the finite mirror aperture,
the cavity can exhibit additional geometrical diffraction losses if there is
misalignment between the two optical axes of the cavity mirrors. This loss
becomes more critical for near-unstable cavities. Hence, we try to assess the
misalignment in our cavity based on the observed variation of cavity linewidth
across the cavity lengths. 
Under the assumption that the misalignment is entirely due to the tilting of the mirrors, the diffraction loss per round trip is given by ~\citep{Hauck1980}:
\begin{equation}
\alpha=\theta^2 \frac{1+g^2}{(1-g^2)^{3/2}} \frac{\pi l_{cav}}{\lambda}\frac{(a/w_{m})^2}{\textrm{exp}[2(a/w_{m})^2]-1}\,,
\end{equation}
where $\theta$ is the misalignment angle, $a$ the radius of cavity mirror
aperture, and $w_{m}$ the beam waist on the mirrors.
Attributing all cavity losses to such diffraction losses bounds
the misalignment to about 0.5 degrees. 
This is compatible with what we expect from the alignment procedure, as the
reflected laser beams from the cavity mirrors are ensured to couple back to
the optical fibers.

\begin{figure}
\centering
  \includegraphics[width=\columnwidth]{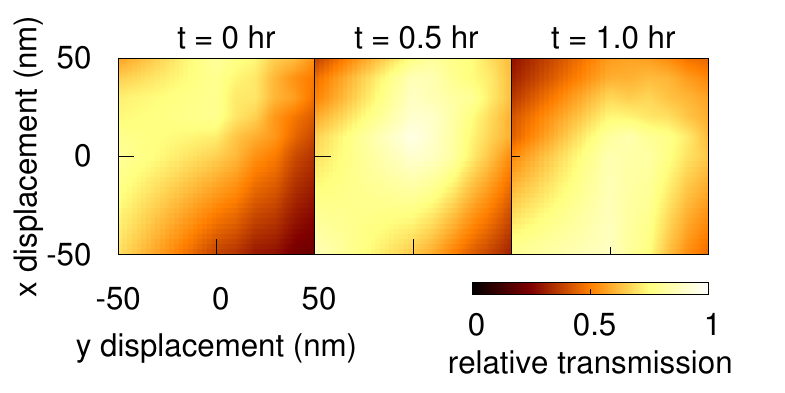}
  \caption{\label{fig:figure5}
    Drift of cavity alignment. Light transmitted through a cavity resonance is
    coupled to a single mode fiber as a mode filter. The relative transmission
    (before and after the single mode fiber) as a function of
    transverse displacement of the second cavity mirror is shown over some
    time without any stabilization.
}
\end{figure}

\section{Transverse stabilization}
The alignment of near-concentric cavities is sensitive to the transverse
positions of the cavity mirrors. To quantify this, we measure the coupling
efficiency of a resonant cavity mode to the mode defined by a single mode
fiber as we displace one of the cavity mirrors in x and y directions (see
Fig.~\ref{fig:figure5}). Throughout the measurement, the cavity length is
locked to the frequency stabilized 810\,nm laser. 
The transmission profiles in Fig.~\ref{fig:figure5} show a FWHM of 42(8)\,nm
and 54(10)\,nm for displacements in x and y.
We observe a drift of 22\,nm of cavity alignment over a time of one hour.
With the thermal expansion coefficients of the cavity setup, such a change of
the fundamental mode transmission by 10\% could be caused by a temperature
change on the order of 100 mK.
Practical operation of a near-concentric cavity therefore requires
either careful temperature stabilization of the setup, or a transverse locking
scheme. 

\begin{figure}
\centering
  \includegraphics[width=\columnwidth]{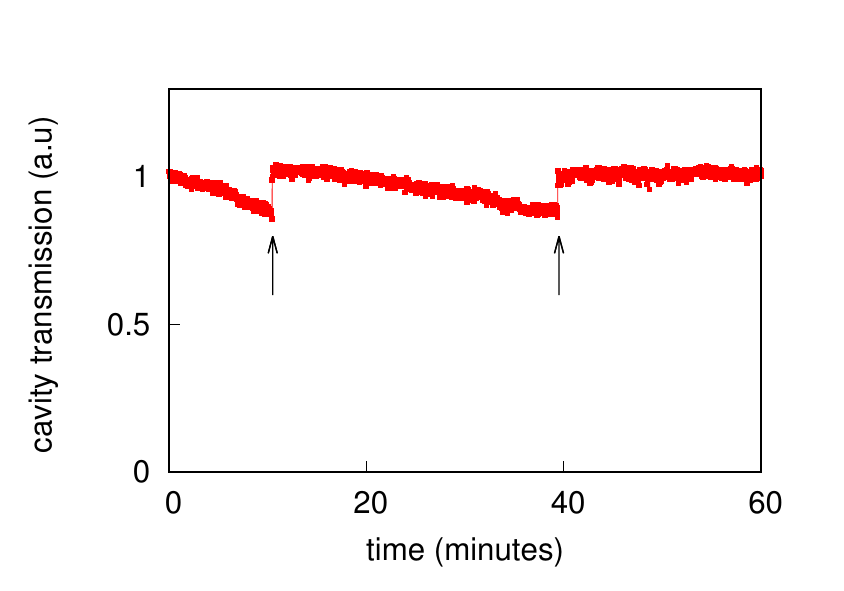}
  \caption{\label{fig:figure6}
Transverse stability of the near-concentric cavity at the last stable
longitudinal resonance ($d=207$\,nm). The slow
drift of cavity transmission on the order of minutes is due to the transverse
misalignment caused by temperature change, while the cavity length is
locked to a probe light resonance during the measurement.
Vertical arrows indicate the activation of the stabilization algorithm, where
the cavity transmission recovers to the maximum value after the successful
implementation of the algorithm within a few seconds. 
}
\end{figure}

 To actively compensate for transverse drifts, we implement a two-dimensional
 lock-in algorithm based on gradient search method to maximize the cavity
 transmission for the two transverse displacement variables.
Figure~\ref{fig:figure6} shows a typical record of cavity transmission at the
last resonant length when the stabilization algorithm is activated at two
instances. The slow drift on the order of minutes between these instances is
probably caused by the temperature change of the cavity.
The average search time to recover the maximum cavity transmission is on the
order of seconds, and thus would not significantly reduce the duty cycle of an
experiment. With both temperature stabilization and active transverse stabilization, the near-concentric cavity remains aligned for a few hours.

\section{Conclusion}
We presented a compact design, alignment procedure and stabilization methods of
a Fabry-Perot near-concentric optical cavity. In our experiment, we find
that the cavity design preserves cavity linewidth and cavity transmission
when being operated at 207(13) nm shorter than the concentric point, the last
longitudinal resonance for this cavity setup.

At this cavity length, the  measured transverse mode frequency spacing of
40(5)\,MHz is of the same order as the estimated atom-cavity coupling strength
of 20\,MHz for a Rb atom placed into the cavity mode.
This permits to probe the dynamics of an atomic state when strongly coupling
to several cavity modes, opening an avenue to experimentally explore
multi-mode cavity QED in the optical regime, and new schemes of interaction of
photons simultaneously present in different modes.

\begin{acknowledgments}This work was supported by the Ministry
  of Education in Singapore and the National Research
  Foundation, Prime Minister’s office (partly under grant no
  NRF-CRP12-2013-03).
\end{acknowledgments}
\bibliographystyle{apsrev4-1}
%\bibliography{cavity}

%merlin.mbs apsrev4-1.bst 2010-07-25 4.21a (PWD, AO, DPC) hacked
%Control: key (0)
%Control: author (72) initials jnrlst
%Control: editor formatted (1) identically to author
%Control: production of article title (-1) disabled
%Control: page (0) single
%Control: year (1) truncated
%Control: production of eprint (0) enabled
%

\end{document}